\documentclass{emulateapj}

\usepackage{graphicx}
\usepackage{float}
\usepackage{amsmath}
\usepackage{epsfig,floatflt}

\begin{document}

\title{A search for concentric circles in the 7-year WMAP temperature sky maps}

\author{I. K. Wehus\altaffilmark{1} and H.\ K.\ Eriksen\altaffilmark{2,3}}

\email{i.k.wehus@fys.uio.no}

\altaffiltext{1}{Department of Physics, University of
  Oslo, P.O.\ Box 1048 Blindern, N-0316 Oslo, Norway}

\altaffiltext{2}{Institute of Theoretical Astrophysics, University of
  Oslo, P.O.\ Box 1029 Blindern, N-0315 Oslo, Norway}

\altaffiltext{3}{Centre of Mathematics for Applications, University of
  Oslo, P.O.\ Box 1053 Blindern, N-0316 Oslo, Norway}

%\date{Received - / Accepted -}

\begin{abstract}
In a recent analysis of the 7-year WMAP temperature sky maps,
Gurzadyan and Penrose claim to find evidence for violent pre-Big Bang
activity in the form of concentric low-variance circles at high
statistical significance. In this paper, we perform an independent
search for such concentric low-variance circles, employing both
$\chi^2$ statistics and matched filters, and compare the results
obtained from the 7-year WMAP temperature sky maps with those obtained
from $\Lambda$CDM simulations. Our main findings are the following: We
do reproduce the claimed ring structures observed in the WMAP data as
presented by Gurzadyan and Penrose, thereby verifying their
computational procedures. However, the results from our simulations do
not agree with those presented by Gurzadyan and Penrose. On the
contrary we obtain a substantially larger variance in our simulations,
to the extent that the observed WMAP sky maps are fully consistent
with the $\Lambda$CDM model as measured by these statistics.
\end{abstract}

\section{Introduction}
\label{sec:introduction}

One of the main accomplishments in cosmology during the last decade is
the establishment of the $\Lambda$CDM inflationary concordance
model. According to this model, the universe consists of 5\% baryonic
matter, 22\% dark matter and 73\% dark energy \citep{jarosik:2010},
and is filled with random and Gaussian fluctuations. These
fluctuations were generated during a short period of exponential
expansion called inflation \citep[e.g.][ and references
  therein]{liddle:2000}, during which the universe expanded by a
factor of $\sim10^{26}$ in $\sim10^{-34}$ s. With only a handful of
free parameters, this model is able to successfully fit thousands of
observational data points.

Nevertheless, the $\Lambda$CDM model must at the current stage be
considered an effective model rather than a fundamental model. First,
it relies on several quantities that have never been directly observed
except through their gravitational impact, such as both dark matter and
dark energy. Second, it postulates the existence of an unknown scalar
field, the inflaton. It is therefore important to put the inflationary
framework to stringent tests, probing its range of validity in
different ways. One approach to do so is to construct alternative
cosmological theories, making different observational predictions than
$\Lambda$CDM, and then compare the two models using high-precision
data.

One example of such work has been demonstrated by the development of
the ``Conformal Cyclic Cosmology'' (CCC) model by
\citet{penrose:2008,penrose:2009,penrose:2010}. In this picture, the
history of the universe is described in terms of a series of
``aeons'', each of which is defined as a finite period between two Big
Bang events. We will not consider further details of the CCC model in
this paper, except for one interesting feature: According to
\citet{gurzadyan:2010}, the model postulates that super-massive black
holes may collide in earlier aeons, releasing tremendous amounts of
energy in the form of gravitational waves. These collisions may be
observable in our current aeon in the form of concentric circles of
low variance in the cosmic microwave background (CMB).

\begin{figure*}[t]
\mbox{\epsfig{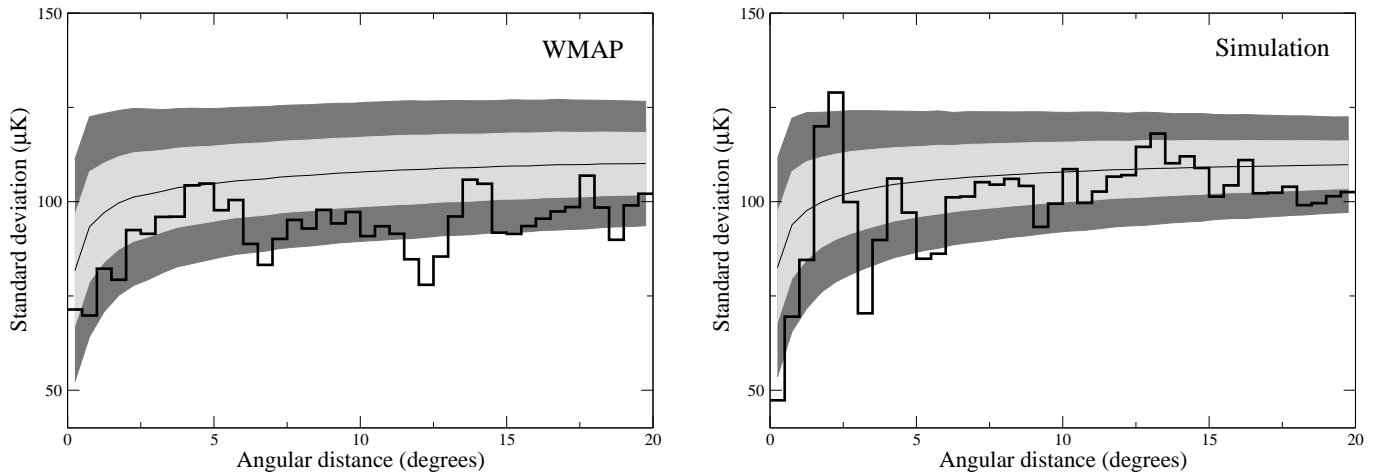}}
\caption{Examples of single standard deviation profiles (thick
  histograms), computed from WMAP (left) and a simulation (right). The
  thin solid line shows the mean of all profiles, and the shaded
  regions show the 1 and $2\sigma$ confidence regions. The WMAP
  profile is centered on galactic coordinates $(l,b) =
  (105.04^{\circ}, 37^{\circ})$, reproducing Figure 2 of
  \citet{gurzadyan:2010}. Note that low-variance rings are found also
  in the simulation.}
\label{fig:examples}
\end{figure*}

Following up on this prediction, \citet{gurzadyan:2010} analyzed the
7-year WMAP temperature sky maps, searching for concentric circles of
low variance. And quite surprisingly, they claim to find evidence for
such rings at the $6\sigma$ confidence level, by comparing the results
obtained from WMAP with $\Lambda$CDM simulations. This claim is
sufficiently spectacular (involving both pre-Big Bang phenomena and
super-massive black holes) to catch the interest of the general media,
with numerous news stories and live media appearances following in the
weeks after the release of the paper. Given these widespread
reactions, the claims of Gurzadyan and Penrose deserve closer scrutiny
through independent analysis; this paper presents one such independent
analysis.

\section{Data and simulations}
\label{sec:data}

The pre-Big Bang black hole collisions discussed by
\citet{gurzadyan:2010} observationally manifest themselves in the form
of one or more rings of low variance centered on a central position,
$\hat{p}$, in the CMB anisotropy field, $\Delta T(\hat{n})$, observed
today. In this paper, we therefore consider the best currently
available full-sky maps of the CMB, namely the 7-year WMAP temperature
sky maps. These data are provided in the form of pixelized HEALPix sky
maps with a pixel resolution of $7'$.

For simplicity we consider only the foreground-reduced WMAP W-band (94
GHz) data in this paper, as this channel has the highest resolution of
all the WMAP bands, with a beam of $13'$ FWHM. (Note that
\citet{gurzadyan:2010} additionally analyses the WMAP V-band data and
the BOOMERanG98 data; however, we agree that instrumental effects are
irrelevant to this analysis, and therefore consider only the W-band in
the following.) We apply the WMAP KQ85 sky cut to this map, removing
22\% of the sky, and taking into account both the galactic plane and
high-latitude point sources \citep{gold:2010}. (Note that
\citet{gurzadyan:2010} adopts a straight $|b| < 20^{\circ}$ sky cut;
the difference is not significant.)

Next, we build an ensemble of 1000 simulations with a spectrum given
by the best-fit WMAP 7-year $\Lambda$CDM model
\citep{komatsu:2010}. These realizations are convolved with the W-band
beam and HEALPix pixel window, and projected onto a HEALPix
$N_{\textrm{side}}=512$ grid. Finally, uncorrelated Gaussian noise
with RMS equal to $\sigma_0/\sqrt{N_{\textrm{obs}}(p)}$ are added to
pixel $p$, where $\sigma_0$ is the W-band noise RMS per observation
and $N_{\textrm{obs}}(p)$ is the number of observations in that
pixel. The KQ85 sky cut is also applied to the simulations.

\section{Method}
\label{sec:algorithms}

The central concept in the following search is the radial standard
deviation profile\footnote{We use the terms ``standard deviation
  profile'' and ``variance profile'' interchangeably in this paper.},
\begin{equation}
\sigma_p(b) \equiv \sqrt{\frac{1}{N-1} \sum_{\theta \in b} \Delta
T(n)^2},
\end{equation}
where $p$ denotes a fixed reference pixel on the sky, $n$ is a
variable pixel index, and $\theta = \textrm{acos} (\hat{p} \cdot
\hat{n})$ is the angular distance between those two positions; $b$
denotes an angular bin, $[\theta_{b-1}, \theta_b)$. Thus, this
  function measures the standard deviation of the CMB field as a
  function of radial distance from $p$, and a sharp negative spike in
  this function corresponds to the postulated signature of the black
  hole collisions. To reduce correlations between different bins, we
  subtract the mean from each function $\sigma_p(b)$ separately
  before further processing.
  
We measure $\sigma_p(b)$ for all positions over a HEALPix
$N_{\textrm{side}}=32$ grid, for a total of 12\,288 points. Each
function is evaluated from 0 to $20^{\circ}$ in 40 bins, corresponding
to a bin size of $0.5^{\circ}$. If more than 40\% of the pixels within
$20^{\circ}$ of a given pixel $p$ are masked out, that pixel is
removed from further analysis. Finally, we also remove any pixels that
are immediate neighbors to the galactic cut.

Having computed these functions for both data and simulations, the
next step is to identify potential low-variance ring candidates, and
quantify their significance. To this end, we introduce two different
statistics. First, we consider matched filters designed to highlight
negative spikes in $\sigma_p(b)$,
\begin{equation}
\hat{\sigma}_p(b) = \sum_{b'} \sigma_p(b') f(b'-b).
\end{equation}
Here $f(b)$ defines a discrete filter, and we consider three different
cases in this paper, namely $f_1 = [0.5, -1, 0.5]$, $f_2 = [0.25,
  0.25, -0.5, -0.5, 0.25, 0.25]$ and $f_3 = [0.5, -0.125, -0.75,
  -0.125, 0.5]$; for $|b-b'|$ larger than the length of the filter,
$f$ is zero. These filters corresponds to 1) a negative top-hat filter
of width 1; 2) a negative top-hat filter of width 2; and 3) a negative
wedge filter, each sensitive to typical interesting candidates. For
each case, we adopt the maximum of $\hat{\sigma}_p(b)$ as our
statistic, considering only angular distances larger than $2^{\circ}$
as the intrinsic estimator variance is very large on the smallest
scales. We then make both sky maps and histograms of the maximum
values of $\hat{\sigma}_p(b)$, and compare these between the observed
data and the simulations.

Our second statistic is based on the standard $\chi^2$ estimator,
which is sensitive to the overall fluctuation level of $\sigma_p(b)$,
rather than individual spikes. Specifically, we compute
\begin{equation}
\chi^2 = \sum_{bb'} (\sigma_{p}(b) - \mu_b) C^{-1}_{bb'}
(\sigma_{p}(b') - \mu_{b'})
\end{equation}
for each pixel and data set, where $\mu_b$ is the mean of
$\sigma_p(b)$ computed from the simulation set, and $C_{bb'}$ is the
covariance matrix. Again, we make sky maps of these values for the
observed data, and compare the observed data with the simulations in
terms of histograms.

\section{Results}
\label{sec:results}

First, we show in the left panel of Figure \ref{fig:examples} the
standard deviation profile computed from the WMAP data centered on
Galactic coordinates $(l,b) = (105.04^{\circ}, 37^{\circ})$. (The mean
is not subtracted in this plot.) This is the same profile as shown in
Figure 2 of \citet{gurzadyan:2010}. The agreement between the two
results is excellent, despite the different masks used by the two
analyses. This validates the routines used to compute the radial
profiles for both analyses.

We have also plotted the mean of the variance profiles computed from
all pixels in Figure \ref{fig:examples}, together with the
corresponding 1 and $2\sigma$ confidence regions. Here we note several
interesting features. First, the WMAP example profile is consistently
low compared to the mean, suggesting strong correlations between
bins. This is typical for all profiles for both WMAP and simulations;
the local variance in a CMB map depends on the scanning strategy of
WMAP, and the entire profile can shift up or down depending on whether
the corresponding pixel is located in the ecliptic plane (high noise)
or in the ecliptic poles (low noise). Since this effect is of little
interest in the search for concentric rings, we subtract the mean
before further analysis to reduce bin-to-bin correlations.

Second, the simulated distribution is very similar to that of WMAP,
having consistent mean and standard deviation. We have also checked
that this distribution is consistent from simulation to simulation.

Third, we see that the lowest point in the WMAP example corresponds to
a $\sim3.3\sigma$ outlier compared to the typical WMAP profile. This
is the point that was claimed to be a $6\sigma$ outlier by
\citet{gurzadyan:2010}. With our simulations, this point appears
considerably less anomalous. Further, it is important to note that
this particular profile is by no means randomly chosen. On the
contrary, it was picked out precisely due to its anomalous behaviour
after searching through more than 10\,000 points. The actual
statistical significance of this particular $3.3\sigma$ outlier is
therefore likely quite low.

Fourth, as illustrated by the example simulation profile in the right
panel of Figure \ref{fig:examples}, it is not difficult to find
individual simulated profiles with ``anomalous'' behaviour. Here one
can also see ``rings'', similar to those observed in WMAP, but created
by chance alone.

To address these issues more rigorously, we are forced to adopt
statistical techniques; visual inspection is not sufficient. We
therefore employ the search algorithms described in Section
\ref{sec:algorithms} to identify peculiar candidates in both the data
and simulations, and consider various statistics based on these
results to interpret significances statistically.

\begin{figure}[t]
\mbox{\epsfig{figure=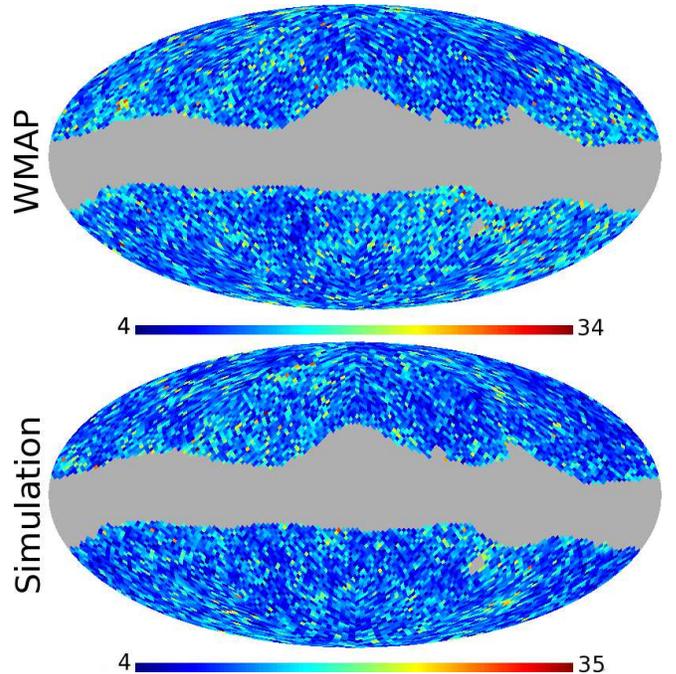,width=\linewidth,clip=}}
\caption{Example of sky maps obtained by the matched filter
  searches, in this case the wedge filter, $f_3$. The top panel shows
  the map obtained from WMAP, the bottom panel shows the same for a
  random simulation.}
\label{fig:skymaps}
\end{figure}

In Figure \ref{fig:skymaps}, we show sky maps of the third matched
filter, $f_3$, statistic, where each pixel indicates the maximum of
$\hat{\sigma}_p(b)$ over $b$. Thus, pixels with large values in these
plots indicate variance profiles with a notable negative wedge-like
structure. Such maps are shown for both the WMAP data and a random
simulation. At least visually, the two maps appear statistically
consistent.

\begin{figure}[t]
\mbox{\epsfig{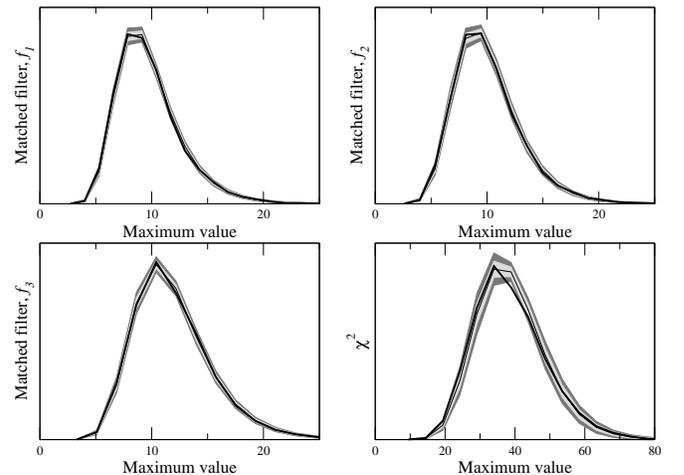}}
\caption{Histogram comparison between the WMAP data (thick solid
  lines) and the $\Lambda$CDM simulations (mean in thin black lines; 1
and $2\sigma$ regions in gray bands), for each of the four statistics
adopted in this paper.}
\label{fig:histograms}
\end{figure}

This statement is quantified more rigorously in Figure
\ref{fig:histograms}, where we show histograms for each of the four
statistics computed from the WMAP data, and compared to the mean
properties of the simulated ensemble. Again, the WMAP properties
appear fully consistent with the simulations.

\begin{figure}[t]
\mbox{\epsfig{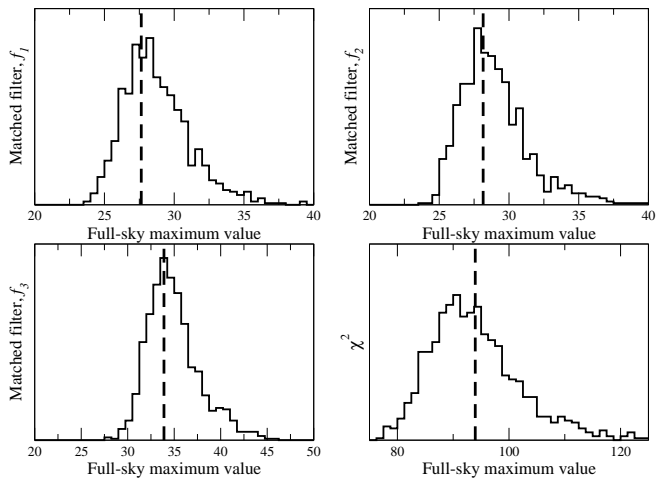}}
\caption{Comparison of the full-sky maximum value for each statistic
  between the simulations (histograms) and the WMAP data (vertical
  dashed line).}
\label{fig:histograms2}
\end{figure}

However, the histogram statistics shown above are sensitive only to
the mean properties of the CMB field. They therefore only show that
there is no evidence for a large number of concentric circles in the
WMAP data, not that there cannot be a small number of highly
significant cases. Finally, we therefore also consider the maximum
value of each of the four statistics, which should be sensitive to
single extreme cases. Figure \ref{fig:histograms2} shows histograms of
such maximum values, as computed from the simulated ensemble, with
the corresponding WMAP value indicated by a vertical line. Again, we
see that WMAP appears fully consistent with the $\Lambda$CDM
simulations. 

\section{Concluding remarks}
\label{sec:conclusions}

In this paper we search for concentric low-variance rings in the
7-year WMAP temperature sky maps, seeking to reproduce the results
recently presented by \citet{gurzadyan:2010}. While our two analyses
do agree in terms of specific variance profiles, they clearly disagree
in terms of statistical interpretation and significance. Specifically,
we find a substantially larger variance in our $\Lambda$CDM
simulations than \citet{gurzadyan:2010} do in theirs. When taking into
account this larger variance, and accounting for statistical selection
effects, the evidence for concentric circles in the WMAP data appears
minimal. Rather, the WMAP data appears fully consistent with our
simulations. 

The main difference between the two analyses must lie in the
construction of the simulations. However, we have good reason to
believe that our simulations are correct. First, we note that in our
simulations the mean variance profile decreases towards small angular
distances. This is bound to happen for two reasons: The CMB
anisotropies constitute a correlated field with a characteristic scale
of $1^{\circ}$, corresponding to the first peak in the CMB
spectrum. Also, the instrumental beam of WMAP smooths out all small
scale structure. It is therefore surprising that this effect is not
seen in Figure 2 of \citet{gurzadyan:2010}. Second, the fact that our
simulations agree with WMAP provides further confidence; it is
difficult to imagine an error in the simulation pipeline that would
cause the simulations to become \emph{more} similar to the observed
data. Third, as noted by \citet{gurzadyan:2010}, there are both low
and high peaks in the reported variance profiles. While Gurzadyan and
Penrose assign no significance to the high variance peaks (stating
that ``the peaks of high variance are of no importance, as these can
result from numerous irrelevant effects''), they are clearly relevant
in our interpretation, as they illustrate the substantial statistical
variance present in these profiles. This large variance is observed
both in the real data and the simulations.

Thus, we conclude that there is no evidence for the CCC model in the
current WMAP data. Of course, Planck \citep{tauber:2010} may provide
new light on this issue, having higher sensitivity and resolution than
WMAP, but one should probably not have too high expectations in this
regard. Even if the CCC model should turn out to describe the real
universe, it will quite likely be difficult to unambiguously identify
such concentric circles due to the dominant background cosmic
variance.

\begin{acknowledgements}
  The computations presented in this paper were carried out on Titan,
  a cluster owned and maintained by the University of Oslo and
  NOTUR. Some of the results in this paper have been derived using the
  HEALPix \citep{gorski:2005} software and analysis package.  We
  acknowledge use of the Legacy Archive for Microwave Background Data
  Analysis (LAMBDA). Support for LAMBDA is provided by the NASA Office
  of Space Science.
\end{acknowledgements}

\end{document}